\documentclass[aps,pra,twocolumn,10pt]{revtex4-1} 
\usepackage[utf8]{inputenc} 
\usepackage{amssymb,amsmath,amsbsy} 
\usepackage{graphicx} 
\usepackage{epsfig} 
\usepackage{setspace}
\usepackage{braket} 
\usepackage{bbm} 
\usepackage{xcolor} 
\usepackage{siunitx} 
\usepackage{times} 
\usepackage{mathtools} 
\usepackage{mathrsfs} 
\RequirePackage{color} 
\RequirePackage[colorlinks=true, urlcolor=blue, anchorcolor=blue, citecolor=blue, filecolor=blue, linkcolor=blue, menucolor=blue, linktocpage=true, pdfproducer=medialab, pdfa=true]{hyperref} 
\usepackage{calligra}  
\DeclareMathAlphabet{\mathcalligra}{T1}{calligra}{m}{n}  
\DeclareFontShape{T1}{calligra}{m}{n}{<->s*[2.2]callig15}{} 
\let\Re\relax 
 
\DeclareMathOperator{\Re}{Re}

\newcommand{\norm}[1]{\left\lVert#1\right\rVert} 
\newcommand{\p}{\partial}

 \DeclareMathOperator{\tr}{tr}

\begin{document} 
\title{Quantum metrology beyond the quantum Cram\'er-Rao theorem} 
\author{Luigi Seveso}
\email{luigi.seveso@unimi.it} 
\affiliation{Quantum Technology Lab, Dipartimento di Fisica dell'Universit\`a degli Studi di Milano, I-20133 Milano, Italia} 
\author{Matteo A. C. Rossi}
\email{matteo.rossi@unimi.it}  
\affiliation{Quantum Technology Lab, Dipartimento di Fisica dell'Universit\`a degli Studi di Milano, I-20133 Milano, Italia} 
\author{Matteo G. A. Paris}
\email{matteo.paris@fisica.unimi.it} 
\affiliation{Quantum Technology Lab, Dipartimento di Fisica dell'Universit\`a degli Studi di Milano, I-20133 Milano, Italia}   
\begin{abstract} 
A usual assumption in quantum estimation is that the unknown parameter labels the possible states of the system, while it influences neither the sample space of outcomes nor the measurement aimed at extracting information on the parameter itself. This assumption is crucial to prove the quantum Cram\'er-Rao theorem and to introduce the quantum Fisher information as an upper bound to the Fisher information of any possible measurement. However, there are relevant estimation problems where this assumption does not hold and an alternative approach should be developed to find the genuine ultimate bound to precision of quantum measurements.  We investigate physical situations where there is an intrinsic dependence of the measurement strategy on the parameter and find that quantum-enhanced measurements may be more precise than previously thought. 
\end{abstract}  
   \keywords{} 
\maketitle 

\section{Introduction}\label{s:intro} 
Basic tasks in quantum metrology fall within the scope of  parameter estimation theory  \cite{helstrom1969quantum,holevo2011probabilistic}, where an  experimenter is interested in learning the value of a parameter  which characterizes the system under study. To this aim, she  performs repeated measurements on the system and then  extracts an estimate of the unknown parameter from the dataset.  The main goal of classical and quantum metrology is that of  optimizing the two following steps: 1) choosing the most suitable  measurement scheme and 2) processing the data to  extract as much information on the parameter as possible.  The typical figure of merit used to assess the performance of  the estimation is the mean square error: the inference strategy  is deemed optimal if the latter achieves a minimum.  
\par 
The second step of the optimization procedure is classical in nature  \cite{Cramer1946mathematical}: once the observable to be measured has  been chosen, the outcomes of the experiment can be treated as the  outcomes of a classical random variable. Following classical  parameter estimation theory, it is well known that for a series  of $n$ independent measurements of a random variable, the minimum  mean square error scales like $1/n$ with a proportionality  coefficient which is equal to the inverse of the  Fisher information (FI). This result goes under the name of  Cram\'er-Rao bound \cite{rao1945information}.   

On the contrary, the first step is the central issue of quantum  metrology \cite{giovannetti2006quantum,paris2009quantum}: setting  aside technological limitations, the experimenter has in principle  the freedom to implement any measurement scheme on the quantum  system. The problem is therefore to select the measurement that  promises the lowest estimation error, as encoded by the corresponding  FI. The FI has been indeed maximized  in the quantum setting over all possible measurement strategies \cite{helstrom1976,braunstein1994statistical},  obtaining the so-called quantum Fisher information (QFI) and  hence the quantum version of the Cram\'er-Rao bound. 
\par 
To introduce the original contribution of this paper, let us emphasize that the maximization in \cite{braunstein1994statistical}  has been obtained assuming that the whole information on the unknown parameter comes from the statistical manifold of possible quantum states of the system. In particular, the measurement strategy aimed at estimating the parameter does not depend on its value. Such an assumption is necessary to the definition of the QFI as the upper bound of the Fisher information corresponding to any possible measurement. More generally, this assumption  is crucial to develop a theory of quantum estimation based on information geometry \cite{brody2013information, amari2007methods,barndorff1986role,BrodyHughston1998, hayashi2006quantum,nagaoka2005fisher,nagaoka2005parameter, Nagaoka2005new}. Indeed, it is only in this case that the QFI simultaneously defines a Riemannian metric on the set of states of the system \cite{ingarden1982information}, and at the same time quantifies the ultimate limits to precision in parameter estimation (thus connecting the statistical properties of the model to the geometrical properties of the state manifold \cite{efron1975defining}). On the other hand, if the statistics of the outcomes captures information on the unknown parameters not only through the state manifold, but also through the details of the measurement strategy, then the link between geometry and statistics is severed. 
\par 
There are many estimation problems where the above assumption does not hold and an alternative approach should be developed to find a proper bound to the ultimate precision allowed by quantum mechanics. 
A first example is represented by  models where there is an additional dependence on the parameter in the measure on the sample space of possible results. This happens e.g. in estimating the algebra deformation  induced by a minimal length at the Planck scale \cite{Rossi2016c}. 
A further example is given by statistical models for Hamiltonian parameters \cite{pang2014quantum,Aharonov2002,Nakayama2015}. In such models a projective measurement on the energy eigenstates intrinsically depends on the unknown parameter. For example, consider one of the standard problems of quantum metrology: magnetometry by employing a spin qubit. The Hamiltonian of the system is given by $H = -\mu_B\, \vec B\cdot \vec \sigma$, where $\vec B$ is the external magnetic field. Its eigenstates do not depend on the magnitude $B$ of the magnetic field, but do depend in a non-trivial way on the polar angle $\theta$ and the azimuthal angle $\phi$ of the qubit with respect to the field's direction. Therefore, the POVM for an energy measurement intrinsically depends on the unknown parameters specifying the direction in space of the spin qubit. This is the particular case we are interested in: the unknown parameter is not a simple phase parameter (the situation most often encountered in a quantum metrological setting), but it appears  in the Hamiltonian in a such a way that its spectrum is parameter-dependent. More generally, parameter-dependent measurements are encountered for observables (e.g. energy, frequency, decay time) which intrinsically depend on the parameter of interest (e.g. coupling constants). 
\par
The paper is organized as follows. In Section \ref{sec:estimation_theory} we set the background and notation for the rest of the paper by introducing classical (\ref{sub:classical_estimation}) and quantum (\ref{sub:quantum_estimation}) estimation theory. We then present the case of a  parameter-dependent sample space in Section \ref{sec:parameter_dependent_sample_space}. In Section \ref{sec:parameter_dependent_povm} we discuss the case of parameter-dependent measurement operators, by analyzing with examples both the case of projective measurements (\ref{sub:projective_povm}) and non-projective measurements (\ref{sub:nonprojective_povm}). We finally discuss the results and their relevance in the framework of quantum estimation theory in Section \ref{sec:conclusions}.

\section{Estimation theory} 
\label{sec:estimation_theory}
\subsection{The classical estimation problem}\label{sub:classical_estimation} 
The classical estimation problem consists in inferring the value of an unknown parameter from $n$ independent measurements of a random variable $X$ whose possible outcomes are subject to statistical uncertainty according to a probability function $p$. 

More formally, the possible outcomes of $X$ have the structure of a probability space $(\chi,\mathcal{B},\mu)$.  The set $\chi$ is called the sample space of $X$, $\mathcal B$ is a $\sigma$-algebra on $\chi$ and $\mu$ is a probability measure. For an event $A\in \mathcal{B}$, $\mu(A)$ is interpreted as the probability of the outcome $A$.  If $\mu$ is absolutely continuous with respect to another measure $\nu$, one may write $\mu(A)=\int_A d\nu\,p$, where $p=d\mu/d\nu$ is a non-negative measurable function (the Radon-Nikodym derivative of $\mu$ with respect to $\nu$). 

Usually, the sample space $\chi$ is taken to be a subset of $\mathbb{R}^n$ and $\nu$ is taken to be the Lebesgue measure on $\mathbb{R}^n$.  However, there are cases in which one has to use the more general measure $\nu$.  Assuming that such measure is absolutely continuous with respect to the Lebesgue measure, one may still write $d\nu= m\,dx$, where $m=d\nu/dx$ is the Radon-Nikodym derivative of $\nu$ with respect to the Lebesgue measure.  Then, the probability functions $\{p_\lambda\}$ are normalized on $\chi$ according to $\int dx\,p_\lambda(x)\,m(x)=1$.  In the classical theory $m$ is assumed to be independent from the parameter $\lambda$, but in the quantum setting it may be necessary  to drop this assumption, as will be discussed later. 
\par 
A statistical model for $X$ consists of a family of probability functions $\{p_\lambda\}$ labeled parametrically by $\lambda\in\Lambda$. The true probability function is obtained for an appropriate choice of $\lambda$, i.e. $p=p_{\lambda^*}$, where $\lambda^*$ is the true value of the parameter. From a geometric  point of view, a statistical model defines a differentiable manifold, with the parametrization $\lambda\to p_\lambda$ providing coordinate charts \cite{shun2012differential}.  

The aim of estimation theory is to produce an estimate for the unknown parameter. To this end one considers an estimator $\hat \lambda$, i.e. a function of $X$ taking values in $\Lambda$. An estimator is said to be unbiased if its expectation value is equal to the parameter $\lambda$, i.e. if $\forall \lambda\in\Lambda$ it holds that $\text{E}_\lambda(\hat \lambda)\equiv\int d\nu\, \hat \lambda(x)\, p_\lambda(x)=\lambda$ (here and in the following, $\text{E}_\lambda$ denotes the expectation value with respect to the element $p_\lambda$ of the statistical model). An estimator is said to be efficient if it minimizes the mean square error with respect to the true value of the parameter. Note that for unbiased estimators the mean square error coincides with the variance $\text{Var}(\hat \lambda)=\text{E}_{\lambda^*}[(\hat\lambda-\lambda^*)^2]$. 

The variance of a general unbiased estimator is bounded from below by the inverse of the Fisher information 
$$\text{Var}(\hat \lambda)\geq[n\,F_X(\lambda^*)]^{-1}$$ 
where the FI is defined as  
\begin{equation}\label{cfisher} 
F_X(\lambda)=\text{E}_\lambda[(\p_\lambda \ln p_\lambda)^2]=\int \!\! d\nu\,[\p_\lambda p_\lambda(x)]^2/p_\lambda(x)\,.
\end{equation} 
An unbiased, efficient estimator is therefore an estimator which achieves equality in the Cram\'er-Rao bound. The  existence of an efficient estimator is guaranteed only for particular statistical models and particular choices of parametrization. However, maximum-likelihood and Bayesian estimators are known to be efficient asymptotically, i.e. when $n\to \infty$ \cite{lehmann1991theory}.    

\subsection{The quantum estimation problem}  
\label{sub:quantum_estimation}
A quantum statistical model is a parametric family of density operators $\{\rho_\lambda\}$ describing  the possible states of a system. The parametrization  $\lambda\to\rho_\lambda$ may be static (e.g.  when $\rho_\lambda$ describes the ground state of a parameter-dependent Hamiltonian) or arise dynamically and may be used to define a differentiable manifold on the set $\{\rho_\lambda\}$, by analogy with the classical case.  The aim of quantum estimation is to produce an estimate of the unknown parameter $\lambda$ from repeated measurements on the system.  

The experimenter has  the freedom to choose among different measurement scheme, formally described by positive operator-valued measures (POVMs) $\Pi$, mapping the $\sigma$-algebra $\mathcal{B}$ of the set $\chi$ of possible outcomes to the space of positive definite, bounded, linear operators on the Hilbert space of the system.  Explicitly, if the outcome of a measurement $x$ belongs to a set $A\in \mathcal{B}$, there is an associated positive operator $\Pi(A)$ such that the probability of such an outcome on the state $\rho_\lambda$ is $\tr(\Pi(A)\rho_\lambda)$. Normalization of probabilities corresponds to the condition  $\int_\chi d\nu\, \Pi(x)={\mathbb I}$  \footnote{In the case of an isolated system, it is enough to consider projective measurements corresponding to some observable, i.e. a  Hermitian operator $X$ with spectral decomposition $X=\sum_{x\in \chi} x\,P(x)$, where $P(x)$ is a projection operator on the subspace spanned by the eigenstates of eigenvalue $x$.}.
\par 
In the quantum setting, an estimator is defined as a couple $(\Pi, \hat\lambda)$ consisting of a POVM $\Pi$ representing the chosen measurement strategy and a classical estimator $\hat \lambda:\chi\to\Lambda$. Its variance is bounded by the inverse of the Fisher information, which is defined in close analogy to Eq. \eqref{cfisher}, i.e. by replacing the probability $p_\lambda(x)$ with its quantum counterpart $\tr(\Pi(x)\rho_\lambda)$, i.e.
\begin{equation}
	F_X(\lambda)=\int\!\! d\nu\,   \frac{[\p_\lambda\tr(\Pi(x)\rho_\lambda)]^2}{\tr(\Pi(x)\rho_\lambda)}.
\end{equation} 
The choice of an estimator $\hat \lambda$ always corresponds to a classical post-processing of the outcomes after the measurement. On the other hand, the choice of the measurement strategy is the central problem of quantum metrology, since different choices in general lead to different attainable precisions. 

Assuming that the unknown parameter only labels the possible states of the system, while it does not influence  the sample space and the POVM, we may follow  \cite{braunstein1994statistical} and maximize the Fisher information over all possible POVMs, obtaining $F_X(\lambda)\leq J(\lambda)$, where the QFI $J(\lambda)$ is defined as 
\begin{equation}   \label{eq:qfi}
J(\lambda)=\text{E}_\lambda(L^2_\lambda)=\tr(\rho_\lambda L^2_\lambda) 
\end{equation}
and $L_\lambda$ is the symmetric logarithmic derivative of $\rho_\lambda$, i.e. $\p_\lambda \rho_\lambda=(\rho_\lambda L_\lambda+L_\lambda \rho_\lambda)/2$.  Therefore, the variance of any estimator $(\Pi,\hat\lambda)$ is bounded by $1/n J(\lambda)$, with $n$ the number of independent measurements. 
The above quantum Cr\'amer-Rao bound found several applications in many different branches of quantum physics, 
where the unknown parameter labels the possible states of the system \cite{rossi15,troiani,Rossi2016c,amelio}.
\par 
The main result of this paper is to point out that the above optimization cannot be used in cases where there is a residual dependence on $\lambda$ apart from the statistical manifold of states. We focus in particular on two scenarios: (i) The measure $\nu$ on the sample space $\chi$ of an observable $X$ depends on $\lambda$. This happens, for example, when the eigenstates of $X$ have a parameter-dependent normalization.  (ii) The POVM depends on $\lambda$. In the projective measurement setting, this may happen because the chosen observable $X$ has eigenstates varying with the unknown parameter. The natural example is an energy measurement since the Hamiltonian $H$ contains the parameter by assumption. In the generalized measurement setting, parameter-dependent POVMs appear for example when the interaction between the system and the meter depends on $\lambda$. In all these cases, the QFI $J(\lambda)$ no longer expresses the ultimate bounds to precision allowed by quantum mechanics.   
Let us now address these scenarios in more detail.

\section{Parameter-dependent sample spaces}\label{sec:parameter_dependent_sample_space} 
Let us assume that the measure $m_\lambda$ on the sample space $\chi$ depends on $\lambda $. As an example, consider the case when we measure a quantum-mechanical observable $X$ with eigenstates $\ket{x}$ normalized as $\braket{x|x'}={m_\lambda(x)}^{-1}\delta(x-x')$.  Then, the completeness relation for the POVM of the projectors on the eigenstates of $X$ is $\int dx\,m_\lambda(x)\,\ket{x}\bra{x}={\mathbb I}$, i.e. the function $m_\lambda$ indeed plays the role of a sample-space measure on $\chi$. {This situation for example is encountered in estimating  the deformation to the commutator between position and momentum induced  by the presence of a minimal length \cite{Rossi2016c}.}
\par 
If the sample-space measure depends on the parameter, a modification of the Cram\'er-Rao bound is necessary already at the classical level \cite{clsup14,clsup16}. The usual Cram\'er-Rao bound of classical statistics is an expression of the Cauchy-Schwarz inequality with respect to the probability inner product $(\cdot\, ,\cdot)$ defined by
\begin{equation}
	(f,g)=\text{E}_\lambda(fg)=\int dx\,m_\lambda(x)\,p_\lambda(x)\,f(x)\,g(x),
\end{equation}
where $f$ and $g$ are any two random variables depending on $X$. In our case, we have to consider the following Cauchy-Schwarz inequality: 
\begin{equation}  |(\hat \lambda-\lambda,\p_\lambda \ln \,m_\lambda p_\lambda)|^2\leq \|\hat \lambda-\lambda\|^2\norm{\p_\lambda\ln \,m_\lambda p_\lambda}^2, \end{equation} where $\|{\hat \lambda-\lambda}\|^2$ is the variance of $\hat\lambda$ and $\norm{\p_\lambda\ln m_\lambda p_\lambda}^2$ is a generalized FI $F^{m_\lambda}_X$, which reduces to the FI of Eq. \eqref{cfisher} when $m_\lambda$ is independent of $\lambda$. The left-hand side can be rewritten as $(\hat \lambda-\lambda, \p_\lambda \ln m_\lambda p_\lambda)=\p_\lambda\text{E}(\hat\lambda)-\lambda\text{E}(\p_\lambda \ln m_\lambda p_\lambda)=1$. The last equality follows from the assumption that the estimator is unbiased and from the fact that the expectation value of the logarithmic derivative of $p_\lambda$ vanishes, since $\text{E}_\lambda(\p_\lambda \ln p_\lambda)=\p_\lambda \int dx\, m(x)\, p_\lambda(x)=0$. 

This leads to an inequality formally identical to the Cram\'er-Rao one, but with the Fisher information redefined as  
\begin{align}   
F^{m_\lambda}_X & \equiv \norm{\p_\lambda\ln m_\lambda p_\lambda}^2   \label{nfinf} \\ 
& = \text{E}_\lambda[(\p_\lambda\ln   p_\lambda)^2]+\text{E}_\lambda[(\p_\lambda \ln m_\lambda)^2] \notag \\ 
& \quad+2\,\text{E}_\lambda(\p_\lambda\ln p_\lambda \, \p_\lambda   \ln m_\lambda). \label{nfinf2}
\end{align} 
Notice that similar inequalities,  though capturing different contributions to the overall fluctuations, have been derived for parameters having themselves an {\em a priori} distribution \cite{vt67} and for (biased) Bayesian estimators \cite{gill95,mang16}. They are often referred to as van Trees  inequalities. 
\par 
Does the parameter-dependence of the sample-space measure always lead to an increase in the available information?  A sufficient condition for the correction to lead to an increase in precision is that $\p_\lambda p_\lambda$ and $\p_\lambda m_\lambda$ have the same sign for all values of $x$, but the answer is in general negative.  

For example, suppose that $m_\lambda$ factorizes as the product of a function of $\lambda$, $m_1(\lambda)$, and a function of $x$, $m_2(x)$: when $\lambda$ is varied infinitesimally, the measure on the sample space at each value of $x$ is simply rescaled by the same factor, independently of $x$.  The vanishing of the expectation value $\text{E}_\lambda(\p_\lambda \ln(m_\lambda p_\lambda))$ immediately implies that $\p_\lambda \ln m_1=-\text{E}_\lambda(\p_\lambda \ln p_\lambda)$ and therefore $$F^{m_\lambda}_X=\text{E}_\lambda[(\p_\lambda\ln p_\lambda)^2]-(\p_\lambda \ln m_1)^2\,.$$
\par 
Going back to the quantum case, from Eq. \eqref{nfinf} it follows that the FI for a measurement of $X$ is given by  
\begin{equation}   
F^{m_\lambda}_X=\int dx\,\frac{[\p_\lambda   (m_\lambda(x)\,\tr(\Pi(x)\rho_\lambda))]^2}{m_\lambda(x)\,   \tr(\Pi(x)\rho_\lambda)}\;.  
\end{equation} 

By taking the derivative with respect to $\lambda$, one finds that 
\begin{equation} 
	\begin{split}\label{nfinfqc} F^{m_\lambda}_X &=\int dx\, m_\lambda(x)\,\tr(\Pi(x)\rho_\lambda)\,(\p_\lambda \ln m_\lambda(x))^2\\&+\int dx\,m_\lambda(x)\,\frac{[\tr(\Pi(x)\p_\lambda\rho_\lambda)]^2}{ \tr(\Pi(x)\rho_\lambda)}\;. 
\end{split}
\end{equation} 
Notice that, since $\int dx\,\p_\lambda m_\lambda\Pi(x)=0$, in the quantum case the double product of Eq. \eqref{nfinf2} identically vanishes $$2\int dx\,\p_\lambda m_\lambda\tr(\Pi(x) \p_\lambda \rho_\lambda)=0\,.$$

Therefore, the Cram\'er-Rao bound for a parameter-dependent measure takes the form $\text{Var}(\hat\lambda)\geq (n\,F^{m_\lambda}_X)^{-1}$, where 
\begin{equation}\label{cr2} 
F^{m_\lambda}_X=\int dx\,m_\lambda(x)\,\frac{[\tr(\Pi(x)\p_\lambda\rho_\lambda)]^2}{\tr(\Pi(x)\rho_\lambda)}+{\mathscr I_m}\;,
\end{equation}
and the information content of the measure  $\mathscr I_m$ is  
\begin{equation}\label{newt}
{\mathscr I_m}=\int dx\, m_\lambda(x)\,\tr(\Pi(x)\rho_\lambda)\,(\p_\lambda \ln m_\lambda(x))^2\;, 
\end{equation} 
which is also the expectation value  $\text{E}_\lambda[(\p_\lambda \ln m_\lambda)^2]$. Thus, in the quantum case the contribution to $F^{m_\lambda}_X$ of the parameter-dependent measure is always positive, i.e. it always leads to an increase in the available information. 

{We remark that the quantum measurement that saturates the QFI $J(\lambda)$ defined in Eq. \eqref{eq:qfi} does not maximize, in general, the FI  $F^{m_\lambda}_X$ of Eq. \eqref{nfinfqc}.  It is not known whether $F^{m_\lambda}_X$ can be maximized in general. Yet,  the first term of $F^{m_\lambda}_X$ can
be maximized following \cite{braunstein1994statistical} and is bounded
by the QFI $J(\lambda)$, so that $F_X\leq J(\lambda)+\mathscr I_m$.
In other words, the Fisher information for a given observable $X$ 
may be larger than the QFI $J(\lambda)$ in the presence of a 
parameter-dependent measure (see also the next Section), though the new bound may not 
be achievable in general.}

\section{Parameter-dependent POVMs}\label{sec:parameter_dependent_povm} Measurements schemes which depend on the unknown parameter can appear naturally in quantum  estimation problems. Nonetheless, the usual derivation of the QFI does not  take into account the possibility of parameter-dependent POVMs.  We can generalize it by considering a measurement with outcomes $\{x\}$ and corresponding POVM $\Pi_\lambda$. 

The Fisher information can be expanded as 
\begin{align}\label{fbound}   F_X(\lambda) ={}&\int d\nu \left\{ \frac{[\tr(\Pi_\lambda(x)\p_\lambda   \rho_\lambda)]^2}{\tr(\Pi_\lambda (x)\rho_\lambda)} +   \frac{[\tr(\p_\lambda \Pi_\lambda(x)\rho_\lambda)]^2}{\tr(\Pi_\lambda   (x)\rho_\lambda)}\right. \notag \\ & \left. +   \frac{2\tr(\Pi_\lambda(x)\p_\lambda \rho_\lambda)\tr(\p_\lambda   \Pi_\lambda(x)\rho_\lambda)}{\tr(\Pi_\lambda (x)\rho_\lambda)}   \right\}. \end{align} 

The first term is the only one that is usually considered and that, when bounded from above, gives the QFI $J(\lambda)$. The remaining two terms are new and should be considered in order to obtain the correct lower bound to the variance of any estimator in the Cramer-Rao theorem. We consider separately first the case of a projective POVM and then of a more general POVM.    

\subsection{Projective POVMs}\label{sub:projective_povm}
Let us focus on the first of the two additional terms appearing in Eq. \eqref{fbound} in the case of a projective measurement of an observable $X$ whose eigenstates $\ket{x}$ depend on $\lambda$. Since $$\tr(\p_\lambda\Pi_\lambda(x)\rho_\lambda)= 2\Re\bra{x}\rho_\lambda\ket{\p_\lambda x}\leq 2|\bra{x}\rho_\lambda\ket{\p_\lambda x}|\,,$$
the term we are interested in can be bounded as 
\begin{equation}\label{bound}   \int d\nu\,\frac{[\tr(\p_\lambda   \Pi_\lambda(x)\rho_\lambda)]^2}{\tr(\Pi_\lambda(x)\rho_\lambda)}\leq   4\int d\nu\,\frac{|\bra{x}\rho_\lambda\ket{\p_\lambda   x}|^2}{\bra{x}\rho_\lambda\ket{x}}\;.\end{equation} 

Let us now define an Hermitian inner product between state vectors of the Hilbert space $\mathcal H$ of the system by $(x|x')_\rho=\bra{x}\rho \ket{x'}$. By using the Cauchy-Schwarz inequality with respect to $(\cdot|\cdot)_\rho$, the term of Eq. \eqref{bound} can be further bounded by
\begin{equation}
	\mathscr{K}_X(\lambda)=4\int d\nu(\p_\lambda x|\p_\lambda   x)_{\rho_\lambda}.
\end{equation} 

From a geometrical point of view, $\mathscr{K}_X$ can be interpreted as follows. Each eigenstate $\ket{x}$ depends parametrically on $\lambda$ and therefore describes a curve in $\mathcal{H}$ with tangent vector  $\ket{\p_\lambda x}$. For a fixed value of $\lambda$, $\mathscr{K}_X$ is the sum of the squares of the lengths of such tangent vectors, which intuitively is a measure of how much the eigenstates of $X$ are sensitive to a variation of the parameter.  Finally, the last term of Eq. \eqref{fbound} can be bounded  by $\sqrt{J(\lambda)}\cdot\sqrt{\mathscr{K}_X(\lambda)}$, again by use of the Cauchy-Schwarz inequality.  In conclusion, for a projective, parameter-dependent measurement of $X$, an upper bound to the FI $F_X(\lambda)$ is given by 
\begin{equation}\label{mcrbound}
F_X(\lambda)\leq (\sqrt{J(\lambda)}+\sqrt{\mathscr{K}_X(\lambda)})^2\;.
\end{equation}  
The quantity on the right has an intuitive appeal. It involves separately the information $J$ coming from the statistical manifold of states and the information $\mathscr K_X$ coming from the parametric dependence of the measurement. When the POVM does not depend on $\lambda$ the usual Braunstein-Caves inequality $F_X\leq J$ is recovered.  

\begin{figure}  \includegraphics[width=0.9\columnwidth]{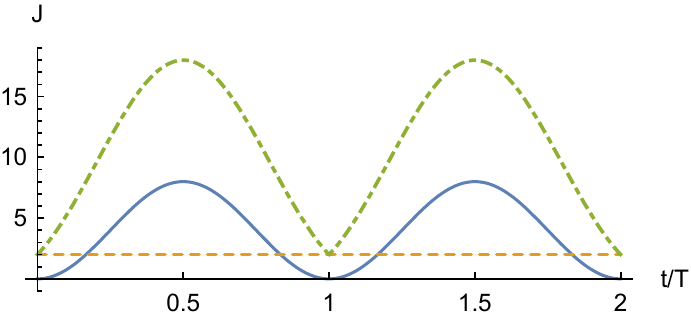}
\caption{\label{comparison}  The QFI $J(g)$ (\emph{solid}), the FI
$F_H(g)$ for an energy measurement (\emph{dashed}) and the upper bound
to the FI, that is the term $(\sqrt{J}+\sqrt{\mathscr{K}_X})^2$ 
of Eq. \eqref{mcrbound}
(\emph{dot-dashed}). As it is apparent from the plot, the QFI $J$ does
not provide the ultimate limits to precision,  since $J<F_H$
periodically with period $T=2\pi/\omega$. For this figure the frequency
$\omega$, the mass $m$ and the displacement $\delta x$ were set to
unity.} \end{figure}  

\subsubsection{Example: quantum gravimetry with a mechanical oscillator} \label{par:example_1}
To give a physical example, we work out the problem of estimating the strength of a uniform gravitational field by employing a mechanical oscillator. The Hamiltonian of the system is 
\begin{equation}
	H=\ p^2/2m +k x^2/2+mg x,
\end{equation} where $m$ is the mass of the oscillator, $k$ its stiffness and $g$ the gravitational field. The  problem is to estimate $g$ through a suitable measurement strategy. To this end, we imagine preparing a coherent state of the oscillator by cooling it to its ground state and henceforth displacing it from equilibrium by $\delta x$. More details, including the wavefunction representation of the coherent state under consideration and its decomposition onto the energy eigenstates, can be found in the Supplementary Material.  
\par
Having delineated both the estimation problem and the statistical manifold, we now investigate the limits to the achievable precision. Among measurement strategies which do not carry any intrinsic information on the unknown parameter, the ultimate bounds are encoded by the QFI $J(g)$, which is given by $J(g)=8m/\omega^3 \sin^2 \omega t/2$, where $\omega=\sqrt{k/m}$ is the oscillator's angular frequency.  
\par 
However, an energy measurement, i.e. a projective measurement on the eigenstates of $H$, depends explicitly on the parameter $g$. Hence the corresponding FI is not guaranteed to be bounded by the QFI just found. Indeed, one can explicitly compute the FI for this particular measurement, which turns out to give $F_H(g)=2 m/\omega^3$. Thus, the Braunstein-Caves inequality $J(g)\geq F_H(g)$ is violated, as can also be seen from Fig. \ref{comparison}. 

In particular, if one is limited by experimental constraints to make very quick measurements, such a strategy can be advantageous. This is due to the fact that, when part of the information on the unknown parameter comes from the POVM, it is unnecessary to wait for the encoding of the parameter onto the manifold of states to happen.  One can also compute the information due to the measurement strategy, $\mathscr K_X$, which turns out to give $\mathscr{K}_X=m/\omega^3 [2+(\xi_\delta -\xi_g)^2]$, where $\xi_\delta$ and $\xi_g$ are defined in Appendix \ref{sec:oscillator}. The upper bound of Eq. \eqref{mcrbound} is also reported in Fig. \ref{comparison}.

As a concluding remark, let us emphasize that, although the projectors on the eigenstates of $H$ explicitly depend on the parameter, the experimenter does not need to know its true value in order to implement the measurement. In other words, an energy measurement has a well-defined meaning which is independent of the particular value of the parameter, even if the measurement outcomes in general depend on it. 

\subsection{Non-projective POVMs}\label{sub:nonprojective_povm}
Let us suppose that the system of interest $S$ is coupled to an ancilla $A$, which plays the role of measuring apparatus. The total Hamiltonian is 
\begin{equation}
	H=H_S \otimes \mathbb{I}_A+\mathbb{I}_S\otimes H_A+H_{SA}=H_0+H_{SA},
\end{equation}
where $\mathbb{I}$ is the identity operator, $H_0$ is the free Hamiltonian and $H_{SA}$ is the interaction Hamiltonian. We are interested in estimating the parameter $\lambda$ which characterizes the dynamics of the system $S$. It is  assumed that the interaction Hamiltonian itself depends on $\lambda$, i.e. both $H_S$ and $H_{SA}$ contain the unknown parameter. 

We will work in the interaction picture, where we are left to solve the Schr\"odinger equation $i\, \partial_t \ket{\psi_{SA}}=H_I\ket{\psi_{SA}}$, where $\ket{\psi_{SA}}$ is the joint state of the system and $H_I=e^{i H_0 t} H_{SA}\,e^{-i H_0 t}$. In general $H_I$ depends  both on $\lambda$ and on  $t$; if, however, $[H_0, H_{SA}]=0$, then $H_I=H_{SA}$. The system is initialized in the product state $\ket{\psi_S}\otimes \ket{0_A}$, where $\ket{0_A}$ is a reference state belonging to the family of eigenstates $\ket{n_A}$ of the ancilla $A$. The evolution operator is the time-ordered exponential $U_t(\lambda)=\mathcal{T} e^{-i\int ^t d\tau\,H_I(\tau, \lambda)}$. 

After an interaction time $T$ a projective measurement on the ancilla is made. The probability of obtaining the eigenstate $\ket{n_A}$ is $p(n)=\tr_{S}(\Pi(n)\,\ket{\psi_S}\bra{\psi_S})$, where $\Pi(n)=M(n)^\dagger M(n)$ and $M(n)= \bra{n_A} U_T(\lambda) \ket{0_A}$.
This is the standard argument of generalized measurements as arising from projective measurements on the ancilla. However, if the interaction Hamiltonian $H_{SA}$ is parameter-dependent then the detection operators $M(n)$ are also parameter-dependent and so is the POVM. This provides a general context in which to look for relevant physical examples. We now discuss a simple one in which a measurement to estimate the frequency of a bosonic mode is performed on a two-level atom interacting with it.

\subsubsection{Example: estimating the frequency of a bosonic mode}\label{par:example_2}

Let us consider a bosonic mode of frequency $\omega$, i.e. $H_S=
\omega(a^\dagger a +1/2)$. {Let us assume that} the quantum
state of the field is confined to the subspace spanned by the vacuum and
the one-photon excitation. The initial state therefore takes the form
$\ket{\psi_S}=c_0 \ket{0}+c_1 \ket{1}$ and the statistical model
consists of the time-evolved states $\ket{\psi_S}_t=c_0 e^{-i\omega t/2}
\ket{0}+c_1 e^{-3i\omega t/2}  \ket{1}$. We are interested in extracting
the value of the parameter $\omega$. To evaluate the ultimate limits to
precision according to the standard theory, we first compute the QFI,
which turns out to give $J(\omega)=4\,t^2 |c_0c_1|^2$. As can be
expected intuitively, it vanishes if the initial state is an energy
eigenstate. 

Now we describe a specific measurement strategy, which consists in coupling the bosonic mode inside a cavity with a two-level atom and measuring whether the atom has remained in its ground state after an interaction time $T$, thus giving rise to a non-projective POVM on the original system. The interaction is described by the Jaynes-Cummings Hamiltonian $H_{SA}= \Omega(a^\dagger \sigma_{-}+a \sigma_{+})$, where $\Omega=d\sqrt{\omega/2\epsilon_0 V}$, $d=\vec {\epsilon}\cdot \bra{e} \vec{d} \ket{g}$, $\vec{\epsilon}$ is the photon polarization, $\epsilon_0$ is the dielectric constant, $V$ the volume of the cavity, $\vec{d}$ the dipole operator, $\ket{g}$ the atom's ground state, $\ket{e}$ the excited state, $\sigma_+=\ket{e}\bra{g}$ and $\sigma_{-}=\ket{g}\bra{e}$. 

We suppose that the atom is in resonance with the radiation field, in which case $[H_0, H_{SA}]=0$, which simplifies somewhat the computations to follow. The system evolves freely for a time $t$ after which it is coupled to the ancilla, i.e. the two-level atom, initialized in its ground state $\ket{g}$. After a time $T$ a projective measurement on the basis $\{\ket{g},\ket{e}\}$ of the atom is made. Since $H_{SA}$ depends explicitly on $\omega$, the resulting POVM is also parameter-dependent, more details are provided in Appendix \ref{sec:jaynes_cummings}. 

One can  compute the FI corresponding to such measurement strategy,
\begin{equation}
F(\omega)=\left(\frac{\Omega T}{\omega}\right)^2\,\frac{|c_1|^2 \cos^2(\Omega T)}{1- |c_1|^2\,\sin^2(\Omega T)}\;.
\end{equation}

{The FI does not vanish in the case $c_0=0$, i.e. if the
bosonic mode is initially in the excited state. Indeed, the scheme just
described allows to extract the information $(\Omega T/\omega)^2$,
whereas standard estimation theory based on computing the QFI suggests
that no information can be extracted on the parameter. The standard
quantum Cramér-Rao bound is violated with this parameter-dependent
POVM.}


\section{Conclusions}\label{sec:conclusions} 
In this paper we have investigated measurements which intrinsically depend on the unknown parameter. Two particular cases have been discussed to illustrate our point. First, a generalized Cram\'er-Rao theorem has been derived, Eq \eqref{cr2}, for a parameter-dependent measure on the space of possible outcomes. The information content due to the sample-space measure takes the form of a new term, Eq. \eqref{newt}, which is additive with respect to the usual FI. Second, the case of a parameter-dependent POVM has been investigated, both in the projective and in the generalized settings. The most interesting result was that the Braunstein-Caves inequality can sometimes be violated, as the FI for a specific parameter-dependent measurement can exceed the QFI. In the projective case, an upper bound on the FI for a parameter-dependent POVM has been derived by introducing the information quantity $\mathscr{K}_X$, which is due to the parametric dependence of the eigenstates of the chosen observable, Eq. \eqref{mcrbound}. Then, a particular example was discussed in Section \ref{par:example_1}. In the generalized case, we argued that parameter-dependent POVMs emerge  when the interaction Hamiltonian between the system and the meter depends itself on the unknown parameter. The example of estimating the frequency of a bosonic mode by use of a two-level atom was developed in Section \ref{par:example_2}. 

In conclusion, our analysis suggests that the standard quantum estimation theory based on the Braunstein-Caves inequality has a more limited scope than previously thought and that implementing measurement strategies that intrinsically depend on the unknown parameter can lead to better precision. Let us also remark that the usual interpretation of the  QFI as the ultimate limit to precision comes from a natural extension of the classical theory. Indeed, both in the classical and quantum cases, the Fisher metric defines a notion of distinguishability on the statistical manifold, which intuitively explains why it can be linked to the precision limit of any estimation strategy. Such a picture changes when the possibility of an intrinsic dependence of the measurement strategy on the parameter is considered, which reflects a difference between the classical and quantum estimation problems. 

\begin{acknowledgments}
We thank Francesco Albarelli and Giacomo Guarnieri for useful discussions. This work has been supported by EU through the collaborative Project QuProCS (Grant Agreement 641277).
\end{acknowledgments}

\bibliography{refncr}

\onecolumngrid
\appendix
\section{Coherent state of a mechanical oscillator}\label{sec:oscillator}
The purpose of this appendix is to provide further information on the coherent state of a mechanical oscillator under gravity, which is the subject of the example discussed in Section \ref{par:example_1}. The Hamiltonian, $H=p^2/2m+kx^2/2+mgx$, has eigenstates 
\begin{equation} \psi_n(\xi)=\left(\frac{m \omega}{\pi}\right)^{1/4}\frac{1}{\sqrt{2^n\,n!}}\,H_n(\xi+\xi_g)\,e^{-(\xi+\xi_g)^2/2}\;,\end{equation}
where $H_n$ is the $n^{\text{th}}$ Hermite polynomial and $\omega=\sqrt{k/m}$. For convenience we introduced the adimensional coordinate $\xi=x/\ell$, with $\ell$ the characteristic length of the oscillator, i.e. $\ell=1/\sqrt{m\omega}$, and in addition defined $\xi_g=mg/k\ell$. We imagine that the oscillator is cooled to its ground state $\psi_0$ and then mechanically displaced from its equilibrium point by $\delta x$. Recall that any coherent state is obtained from the vacuum through the action of the displacement operator, i.e. $\ket{\alpha}=D(\alpha)\ket{0}$, where $D(\alpha)=\text{exp}(\alpha a^\dagger - \alpha^* a)$. If $\delta x$ is the displacement away from the origin imparted to the mechanical oscillator at $t=0$, then our state corresponds to the particular choice $\alpha=-(mk/4)^{1/4}\,\delta x$. The initial preparation is therefore given by
\begin{equation} \psi(x,0)=\left(\frac{m \omega}{\pi}\right)^{1/4}\,e^{-(\xi+\xi_\delta)^2/2}\;,\end{equation} 
where $\xi_\delta=\delta x/\ell$. Its time evolution is described by the family of wavefunctions, parametrized by $g$,
\begin{equation} \psi(x,t)=\left(\frac{m \omega}{\pi}\right)^{1/4}\, e^{-i\omega t(1- \xi_{g}^2)/2}\, e^{-(\xi+\xi_{g})^2/2}\,\text{exp}\left[-e^{-i \omega t}\left(\frac{(\xi_\delta-\xi_g)^2}{2}\cos{\omega t} +(\xi_\delta-\xi_g)(\xi+\xi_g)\right)\right]\;,\end{equation}
which represent the statistical model under study. Its decomposition into energy eigenstates takes the form 
\begin{equation} \psi(x,t)=\sum_{n=0}^\infty c_n\,\psi_n(x)\,e^{-iE_n t}\;,\end{equation}
where the energy spectrum is $E_n=\omega (n+1/2) -mg^2/2\omega^2$ and the coefficients $c_n$ can be computed to be
\begin{equation} c_n=\frac{(-1)^n}{\sqrt{2^n n!}}\,(\xi_\delta -\xi_g)^{n}\,e^{-(\xi_\delta-\xi_g)^2/4}\;.
\end{equation}
The computation of the QFI and FI for an energy measurement proceeds straightforwardly and the results are reported in the main text.  The computation of the corrected bound requires to make use of some integral identities involving products of Hermite polynomials, which we report here for completeness. Let us define the family of integrals
\begin{equation}
 I_p^{n,m}=\int_{-\infty}^\infty d\xi\,\xi^p H_n(\xi)\,H_m(\xi)\,e^{-\xi^2}\;.
\end{equation}  
By integration by part, one can prove the following recurrence relation
\begin{align}\label{rrel}\notag
  I_p^{n,m} & =-\frac{1}{2}\,\int_{-\infty}^\infty d\xi\,\xi^{p-1}\,H_n(\xi)\,H_m(\xi)\,\partial_\xi e^{-\xi^2} =\frac{1}{2}\,\int_{-\infty}^\infty d\xi\,\partial_\xi (\xi^{p-1}\,H_n(\xi)\,H_m(\xi)) e^{-\xi^2} \\ 
  & = \frac{p-1}{2}\,I_{p-2}^{n,m}+n\,I_{p-1}^{n-1,m}+m\,I_{p-1}^{n,m-1}\;.
\end{align}
Starting from the normalization condition for the Hermite polynomials, which corresponds to setting $p$ to $0$,
\begin{equation}
  I_0^{n,m}= \sqrt \pi\, 2^n n!\,\delta_{nm}\;,
\end{equation}
one can use the recurrence relation to find a closed-form expression of $I_p^{n,m}$ for all $p$. For example,
\begin{gather}
  I_1^{n,m}=\sqrt{\pi}\,2^{n-1}\,n!\,\delta_{n-1,m}+\sqrt \pi\,2^n\,(n+1)!\,\delta_{n+1,m}\,,\notag\\ 
   I_2^{n,m}=\sqrt{\pi}\,2^{n-1}\,n!\,(1+2n)\,\delta_{nm}+\sqrt\pi\, 2^{n-2}\,n!\,\delta_{n-2,m}+\sqrt\pi\, 2^n\,(n+2)!\,\delta_{n+2,m}\;,
\end{gather}
and so on. 

\section{POVM for frequency estimation in the Jaynes-Cummings model}\label{sec:jaynes_cummings}

In this appendix, the explicit form of the POVM for the example discussed in Section \ref{par:example_2} is provided. First of all, the evolution operator $U_T(\omega)$ can be checked to be given by
\begin{equation}
  U_T(\omega)=\cos(\Omega T\sqrt{N})\ket{g}\bra{g}+\cos(\Omega T\sqrt{1+N})\ket{e}\bra{e}-i\frac{\sin(\Omega T\sqrt{N})}{\sqrt{N}} a^\dagger\ket{g}\bra{e}-i\frac{\sin(\Omega T\sqrt{1+N})}{\sqrt{1+N}} a\ket{e}\bra{g}\;,
\end{equation}
where $N=a^\dagger a$ is the number operator for the radiation field. The detection operators are
\begin{equation}
  M(g)=\bra{g}U_T(\omega)\ket{g}=\cos(\Omega T \sqrt{N})\;,\qquad\qquad M(e)=\bra{e}U_T(\omega)\ket{g}=-i\frac{\sin(\Omega T\sqrt{1+N})}{\sqrt{1+N}} a\;.
\end{equation}
The corresponding POVM is therefore
\begin{equation}
  \Pi(g)=M(g)^\dagger M(g)= \cos^2(\Omega T \sqrt{N})\;,\qquad \qquad \Pi(e)=M(e)^\dagger M(e)=\sin^2(\Omega T \sqrt{N})\;,
\end{equation}
which depends on the parameter $\omega$ through $\Omega$. 

\end{document}